\documentclass[aps,prl,twocolumn,superscriptaddress,showpacs]{revtex4}
\usepackage{graphicx}
\usepackage{latexsym}
\usepackage{amssymb}
\usepackage{amsmath}
\usepackage{amsfonts}
\usepackage{bm}
\usepackage{multirow}
\usepackage{color}
\newcommand{\ii}{\mathrm{i}}

\newcommand{\cB}{\mathcal{B}}
\newcommand{\cA}{\mathcal{A}}

\renewcommand{\Re}{\mathop{\mathrm{Re}}}
\renewcommand{\Im}{\mathop{\mathrm{Im}}}

\newcommand{\U}{\mathrm{U}}

\newcommand{\beq}{\begin{equation}}
\newcommand{\eeq}{\end{equation}}
\newcommand{\beqn}{\begin{eqnarray}}
\newcommand{\eeqn}{\end{eqnarray}}

\DeclareMathAlphabet{\mathbbold}{U}{bbold}{m}{n}

\def\U{{\rm U}}

\newcommand{\dtriangle}{\bigtriangledown}
\usepackage{ulem}
\usepackage[colorlinks=true,citecolor=blue,linkcolor=blue]{hyperref}

\begin{document}

\title{Pascal's Triangle Fractal Symmetries}

\author{Nayan E. Myerson-Jain}

\affiliation{Department of Physics, University of California,
Santa Barbara, CA 93106, USA}

\author{Shang Liu}

\affiliation{Kavli Institute for Theoretical Physics, University
of California, Santa Barbara, CA 93106, USA}

\author{Wenjie Ji}

\affiliation{Department of Physics, University of California,
Santa Barbara, CA 93106, USA}

\author{Cenke Xu}

\affiliation{Department of Physics, University of California,
Santa Barbara, CA 93106, USA}

\author{Sagar Vijay}

\affiliation{Department of Physics, University of California,
Santa Barbara, CA 93106, USA}

\begin{abstract}
We introduce a model of interacting bosons exhibiting an infinite
collection of fractal symmetries -- termed ``Pascal's triangle
symmetries" -- which provides a natural $\U(1)$ generalization of
a spin-(1/2) system with Sierpinski triangle fractal symmetries
introduced in Ref. \cite{NM}. The Pascal's triangle symmetry gives
rise to exact degeneracies, as well as a manifold of low-energy
states which are absent in the Sierpinski triangle model. Breaking
the $\U(1)$ symmetry of this model to $Z_p$, with prime integer
$p$, yields a lattice model with a unique fractal symmetry which
is generated by an operator supported on a fractal subsystem with
Hausdorff dimension $d_H = \ln (p(p+1)/2)/\ln p$. The Hausdorff
dimension of the fractal can be probed through correlation
functions at finite temperature. The phase diagram of these models
at zero temperature in the presence of quantum fluctuations, as
well as the potential physical construction of the $\U(1)$ model
are discussed.

\end{abstract}

\maketitle

{\it Introduction:} In recent years, generalizations of the notion
of symmetry have significantly broadened our view on states of
matter. Many previously developed phases, such as $Z_2$
topological order~\cite{subirz2,wenz2,ms2001,kitaevz2}, and the
$(3+1)d$ algebraic spin liquid with photon
excitations~\cite{wen2003,ms2003,hermele2004}, were thought to be
beyond the notion of spontaneous symmetry breaking (SSB) of
ordinary global symmetries. But it has been realized in recent
years that these phases still have a unified description as the
SSB of generalized higher-form
symmetries~\cite{formsym0,formsym1,formsym2,formsym3,formsym4,formsym5,formsym6,formsym7,Cordova2019}.
The notion of subsystem symmetry has further enriched our
understanding along this line. Long-range-entangled quantum phases
with fractionalized excitations that exhibit restricted mobility
-- termed fracton orders -- exhibit emergent subsystem symmetries
~\cite{Sagarduality,Sagarclassification, Haahcode,
fractonreview1,fractonreview2}. Subsystem symmetries can be
categorized into type-I and II \cite{Sagarduality}, where a type-I
subsystem symmetry has generators and conserved charges defined on
regular submanifolds of the system, such as lines and planes,
while type-II subsystem symmetries have conserved charges defined
on a fractal-shaped subsystem \cite{yoshida,Haahcode}, often with
non-integer spatial dimensions.

The simplest model with a fractal subsystem symmetry is the
Sierpinski-triangle model, which was first introduced for the
purpose of studying glassy dynamics in the absence of
disorder~\cite{NM}: \begin{equation} H_{\mathrm{ST}} =
\sum_{\bigtriangledown}-K  \sigma^z_1 \sigma^z_2 \sigma^z_3.
\label{st} \end{equation} Here $\sigma^z_i$ is an Ising spin
defined on each site of a triangular lattice, and the sum is only
over the downward-facing triangular plaquettes of the lattice.
This simple model has the following desirable features. (1) The
model has an exotic ``fractal symmetry", which becomes most
explicit when the system is defined on a $L \times L$ lattice with
$L = 2^k - 1$: the Hamiltonian is invariant under flipping spins
on a Sierpinski-triangle fractal subsystem; (2) at finite
temperature, the three point correlation function $\langle
\sigma^z_{0,0}\sigma^z_{r,0}\sigma^z_{0,r} \rangle$ is nonzero
only when $r = 2^k$, and scales as $\sim \exp( - r^{d_H})$ with
$d_H = \ln 3/ \ln 2$, which is the Hausdorff dimension of the
Sierpinski-triangle~\cite{yoshida}; (3) with a transverse field
$\sum_i - h \sigma^x_i$, the system becomes a quantum
Sierpinski-triangle model, and at $h = K$ there is a quantum phase
transition at zero temperature~\cite{juan1,frankyizhi}, which
separates the ``fractal-ordered" phase that spontaneously breaks
the fractal symmetry ($K > h$), and a disordered phase with ($h >
K$).

\begin{figure}
\includegraphics[width=0.45\textwidth]{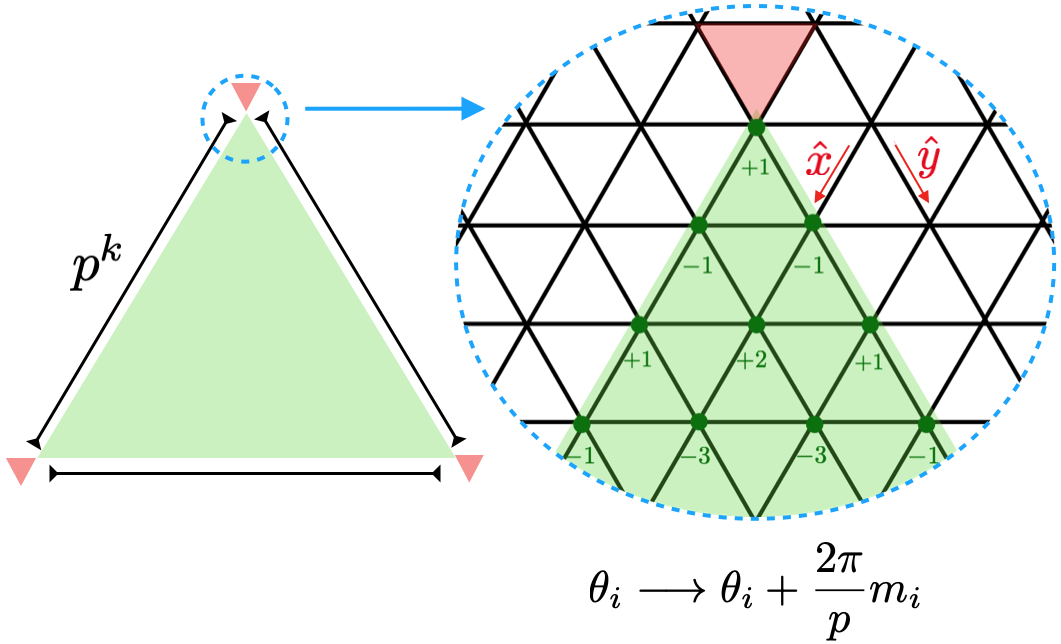}
\caption{{\bf Pascal's Triangle Symmetries:} The $\U(1)$ parent
model (\ref{eq:three_boson}) has a family of fractal symmetries,
generated by a staggered rotation of the boson phase $\theta_{i}
\rightarrow \theta_{i} + \frac{2\pi}{p}m_{i}$ over a triangular
region of side length which is a power of any prime number $p$.
{For each $p$, the fractal symmetry becomes exact for system size
$L^2$ with $L = p^k - 1$.} When acting on a classical ground-state
of the parent model, this transformation generates  excitations at
the corners of the triangular region. The non-trivial action of
this rotation can be visualized as Pascal's triangle modulo $p$,
which is a fractal with Hausdorff dimension $d_H(p) = {\ln
(p(p+1)/2)}/{\ln p}$. } \label{q2fractal}
\end{figure}

In this work we introduce generalizations of both the classical
and quantum Sierpinski-triangle models. These models are obtained
from a $\U(1)$ parent model with ``Pascal's triangle" (also called
Yang Hui triangle in China) symmetries, a family of symmetry
transformations along a fractal region which are exact in a system
with periodic boundary conditions, and for particular system
sizes. Even when these symmetries are not exact, the presence of
an ``approximate" Pascal's triangle symmetry  gives rise to
low-energy states which are absent in the Sierpinski triangle
model \cite{NM}. Descendant models are  obtained by reducing the
$\U(1)$ degree of freedom of the parent model to $Z_p$, with prime
integer $p$. The $Z_p$ models have their own fractal symmetry that
are deduced from the Pascal's triangle symmetry, and their
degenerate excitations have an emergent fractal structure with
Hausdorff dimension $d_H = \ln (p(p+1)/2) / \ln p$.

{\it The $\U(1)$ parent model:} The Hamiltonian of the $\U(1)$
generalization of the Sierpinski-triangle model reads  \beqn
\label{eq:three_boson} H_{\U(1)} = \sum_{\bigtriangledown} - t
\cos\left( \theta_{1} + \theta_{2} + \theta_{3}\right).
\label{U(1)}\eeqn It is straightforward to see that Eq.~\ref{U(1)}
has two conventional $\U(1)$ global symmetries: \beqn \U(1)_1 :
\theta_{j \in A} \rightarrow \theta_{j \in A} + \alpha, \ \
\theta_{j \in B} \rightarrow \theta_{j \in B} - \alpha, \cr\cr
\U(1)_2 : \theta_{j \in A} \rightarrow \theta_{j \in A} + \beta, \
\ \theta_{j \in C} \rightarrow \theta_{j \in C} - \beta.
\label{2U(1)} \eeqn $A, B, C$ are the three sublattices of the
triangular lattice. The ground states of Eq.~\ref{U(1)}
spontaneously break the two $\U(1)$ symmetries. Starting with one
of the ground states, say $\theta = 0$ uniformly on the entire
lattice,  a class of ground states can be generated by rotating
$\theta$ globally according to Eq.~\ref{2U(1)}. Any ground state
obtained this way still has a uniform order of $\theta$ on each of
the three sublattices, hence the ground states generated through
Eq.~\ref{2U(1)} have a conventional ``$\sqrt{3} \times \sqrt{3}$"
order, which is the order often observed on the triangular lattice
antiferromagnet.

Besides the two ordinary $\U(1)$ global symmetries, this model
Eq.~\ref{U(1)} actually contains an infinite series of $Z_p$
distinct fractal symmetries, one for each prime number $p$, $p
\geq 2$. In contrast to the classical Sierpinski-triangle model,
where the $Z_2$ fractal subsystem symmetry is taken as a sequence
of spin flips in the shape of a Sierpinski-triangle, the series of
$Z_p$ fractal symmetries exhibited by the $\U(1)$ parent model are
in the shape of a Pascal's triangle modulo $p$. For example when
$p = 2$, this Pascal's triangle symmetry reduces down to the
familiar $Z_2$ fractal symmetry of the Sierpinski-triangle model;
for $p=3$ the Pascal's triangle modulo 3 reduces to another
fractal shape (Fig.~\ref{q2fractal}).

The exact series of fractal  transformations of Eq.~\ref{U(1)} can
be written down as a staggered rotation of the $\theta_i$'s in the
shape of a Pascal's triangle modulo $p$, which has a side-length
of $p^{k}-1$, where $k$ is any integer greater than zero. The
precise form of the  transformation is \beqn \theta_i
\longrightarrow \theta_i + \frac{2 \pi}{p} (-1)^{i_x+i_y} {i_x+i_y
\choose i_y}  .\label{Frtransform} \eeqn at the points
$(i_{x},i_{y})$ for which $0\le i_{y}\le i_{x}$ and $i_{x}+i_{y}
\in[0,p^{k}-1]$.  As shown in the Supplemental Material (SM),
transformations of this form can be used to generate exact
symmetries when the system is placed on an $L\times L$ lattice
with periodic boundary conditions and with  $L = p^{k}-1$.

Any $Z_p$ fractal transformation of the $\U(1)$ parent model which
is not an exact symmetry of the model generates fully immobile
defects which are analogous to fractons. From the uniform
$\theta_i = 0$ ground-state, transforming the $\U(1)$ degrees of
freedom according to Eq.~\ref{Frtransform} in the shape of a local
Pascal's triangle of size $p^k-1$ creates three defects of energy
$t(1 - \cos({2 \pi }/{p} ))$, one at each downward-facing
triangular plaquette located at the corners of the Pascal's
triangle, as shown in the SM, and as indicated schematically in
Fig.~\ref{q2fractal}. If we treat these defects as point-like
quasiparticles localized on their downward facing plaquettes,
individual defects cannot be moved by any rotation of $\theta_i$'s
without creating more excitations and are hence completely
immobile. We note that since $p$ can be arbitrarily large, the
excitations created by a $Z_{p}$ fractal transformation can cost a
vanishingly small energy in a thermodynamically large system.

\begin{figure}
\begin{center}
\includegraphics[width=0.4\textwidth]{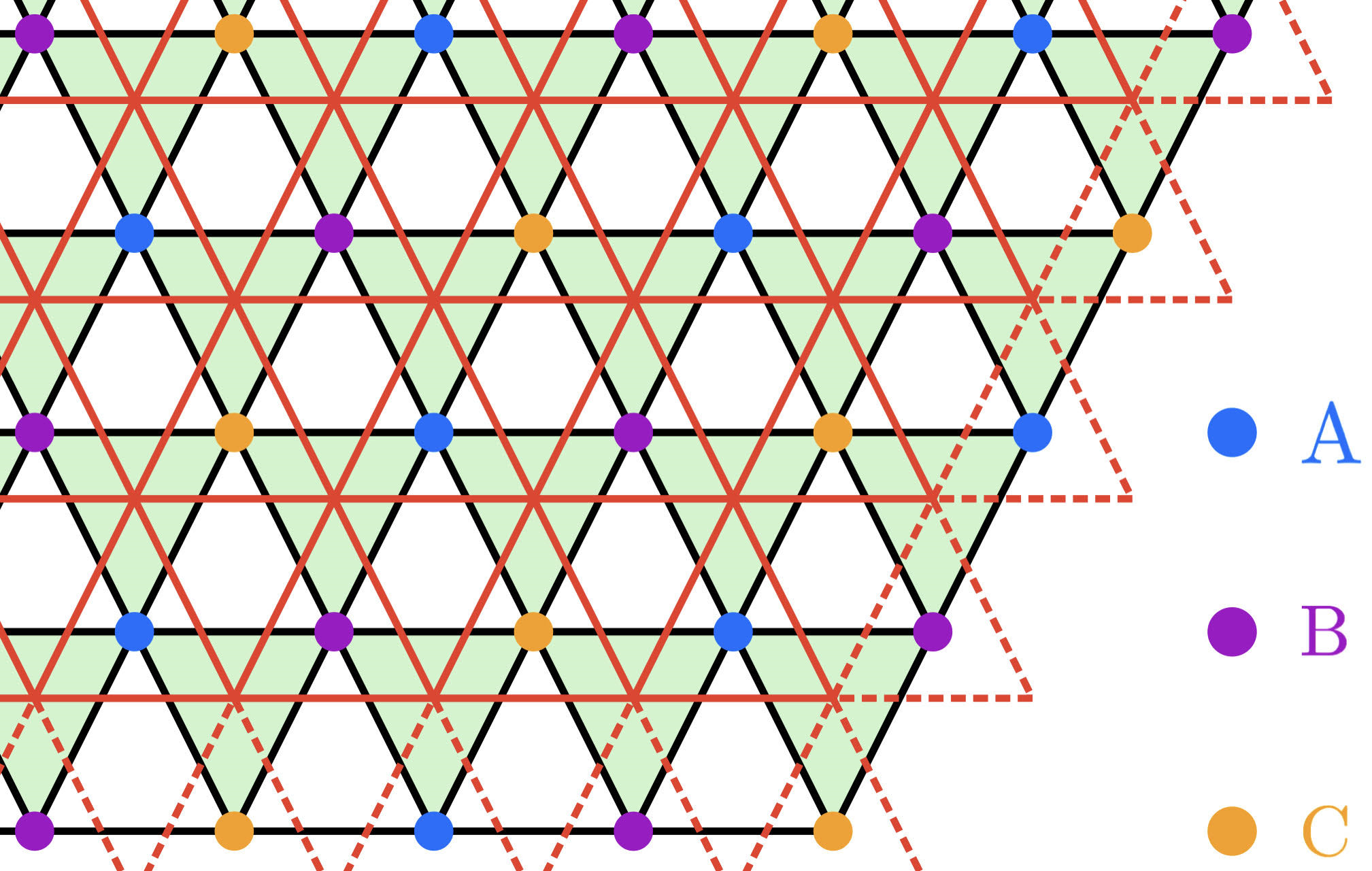}
\caption{{\bf The U(1) Model and its Dual Description:} The models
we consider in this work involve sum of all the downward facing
triangles (shaded in green); The dual of the $\U(1)$ model
(Eq.~\ref{dualc}, Eq.~\ref{dual}) is defined on the dual
triangular lattice, whose sites are the center of each downward
facing triangles of the original lattice.}\label{dualtriangle}
\end{center}
\end{figure}

At finite temperature, the $\U(1)$ parent model is completely
disordered similarly to the Sierpinski-triangle model~\cite{NM}.
This can be most easily seen from a duality mapping of the $\U(1)$
degrees of freedom on the vertices to new $\U(1)$ degrees of
freedom on downward facing plaquettes $(\theta_1 +
\theta_2+\theta_3)_\bigtriangledown \rightarrow \phi_\dtriangle$,
where $\phi_\dtriangle$ is defined on the dual site located at the
center of each downward facing triangular plaquette
(Fig.~\ref{dualtriangle}), and $\phi$ is still compact
(periodically defined). The dual of the $\U(1)$ parent model is
\beqn H_{\U(1)}^d = \sum_\dtriangle -t \cos(\phi_\dtriangle).
\label{dualc}\eeqn Since each $\phi$ is decoupled from one
another, the partition function factorizes into a product of local
partition functions for each individual $\phi$ which does not
support any phase transition.

The three-body interactions of the Sierpinski-triangle model, as
well as the $\U(1)$ parent model look artificial.
Ref.~\cite{nayanryd} proposed to realize the Sierpinski-triangle
model with the Rydberg atoms with only two-body Van der Waals
interactions. In the SM we present a more natural construction of
the $\U(1)$ parent model through a set-up with only two-body
interactions.

{\it The $Z_p$ models:}  From the $\U(1)$ parent model
Eq.~$\ref{U(1)}$, models with a single fractal symmetry that are
natural extensions of the Sierpinski-triangle model can be
constructed. This is done by breaking the $\U(1)$ degrees of
freedom down to $Z_p$ clock degrees of freedom $\sigma_i = e^{i
\theta_i}, \theta_i \in \frac{2 \pi}{p} Z_p$: \beqn H_{Z_p} &=&
\sum_\bigtriangledown -\frac{t}{2} \sigma_1 \sigma_2 \sigma_3 +
\text{h.c.} \label{Zp} \eeqn

The model Eq.~\ref{Zp} extends many properties of the
Sierpinski-triangle model to a more general series of $Z_p$
``Pascal's triangle models'' which reduces to Eq.~\ref{st} when $p
= 2$. Generally, since the $Z_p$ fractal symmetry of the Pascal's
triangle models are descended from the $\U(1)$ parent model, the
fractal symmetry transformation in these models is realized by
Eq.~\ref{Frtransform} with the appropriate choice of $p$. These
series of models also display the fracton-like defects associated
to fractal excitations in the shape of a Pascal's triangle modulo
$p$ that cost energy $t(1 - \cos ( \frac{2 \pi}{p} ))$ each as
well as spontaneously breaking the $Z_p$ fractal symmetry,
yielding a ground-state degeneracy of $p^{L-1}$ when $L = p^k-1$
(see SM for derivation).

Spontaneous symmetry breaking of the $Z_p$ fractal symmetries in
the Pascal's triangle models can be diagnosed by the behavior of a
three-point correlation function. Making use of the duality of
these models, we can define plaquette degrees of freedom
$\tau_\dtriangle = (\sigma_1 \sigma_2 \sigma_3)_{\dtriangle}$. The
dual Hamiltonian is \beqn H_{Z_p}^d = \sum_\dtriangle -
\frac{t}{2} \tau_\dtriangle + \text{h.c.} \label{Zpd} \eeqn In the
thermodynamic limit, each $\sigma$-variable can be represented as
an infinite staggered product of dual $\tau$-variables in the
shape of a Pascal's triangle modulo $p$. The three-point function
$ \mathcal{C}_3(r) = \langle \sigma_{0,0} \sigma_{r,0}
\sigma_{0,r} \rangle$,  after being rewritten in terms of the dual
variables only has compact support when $r = p^k$ and hence must
vanish else wise. The three-point function factors into a product
of single-site expectation values $\langle \tau \rangle, \dots,
\langle \tau^{p-1} \rangle$. From the form of the Hamiltonian, a
general expression for the three-point function for arbitrary $p$
prime can be derived (see SM for details) \beqn \mathcal{C}_3(r =
p^k) = \prod_{m=1}^{\frac{p-1}{2}} \langle \tau^m
\rangle^{N_{m,p-m}(k)} \label{3point}, \eeqn $N_{m,p-m}(k)$ is the
number of times $m$ and $p-m$ appear in a Pascal's triangle modulo
$p$ with length $p^k-1$. Such an expression is complex, but a
complete set of recurrence relations is constructed in the SM for
$N_{m,p-m}(k)$ lending Eq.~$\ref{3point}$ to efficient numerical
evaluation. For small $p$ this can be done analytically, e.g. for
$p = 3$, the three-point function is \beqn && \mathcal{C}_3(r =
3^k) = \langle \tau \rangle^{r^{d_H}} \cr\cr &=& \bigg (
\frac{e^{\beta t} - e^{- \beta t/2}}{e^{\beta t}+2e^{- \beta t/2}}
\bigg )^{r^{d_H}} \label{3pointZ3}  \sim e^{-\alpha r^{d_H}},
\eeqn where $d_H = \frac{\ln 6}{\ln 3}$ is the Hausdorff dimension
of a Pascal's triangle modulo $3$. In general there are natural
bounds on the decay of the three-point function: $\mathcal{C}_3(r
= p^k)$ will decay hyper-exponentially as $\sim e^{-\alpha
r^{\gamma}}$ for sufficiently large $r$, where the exponent of the
decay $\gamma$ is bounded such that \beqn d_H \leq \gamma \leq
\bigg (\frac{p-1}{2} \bigg ) d_H, \text{ } p > 2 \label{hyperexp}.
\eeqn

As demonstrated by Eqs.~\ref{3point} and \ref{hyperexp}, the decay
of the three-point function can be complicated for general $p$
prime. However, a modified version of the $Z_p$ Pascal's triangle
models in Eq.~\ref{Zp} can be proposed for which the three-point
function at finite temperature always decays as a fractal
area-law. If we consider an equal-weight summation of plaquette
terms \beqn \mathcal{H}_{\text{p}} = \sum_{\dtriangle}
\sum_{m=0}^{\frac{p-1}{2}} - \frac{t}{2} (\sigma_1 \sigma_2
\sigma_3)^m + \text{h.c.} \eeqn This model retains the $Z_p$
fractal symmetry of Eq.~\ref{Zp} as it only includes products of
the original plaquette terms. As such, the duality $(\sigma_1
\sigma_2 \sigma_3)_\dtriangle \rightarrow \tau_\dtriangle$ still
exists and the dual of $\mathcal{H}_p$ is \beqn
\mathcal{H}_{\text{p}}^d = \sum_{\dtriangle}
\sum_{m=0}^{\frac{p-1}{2}} - \frac{t}{2} \tau_\dtriangle^m +
\text{h.c.} \eeqn The manner in which the three-point correlation
for $\mathcal{H}_{\text{p}}$ is calculated remains the same as for
what it was in Eq.~$\ref{Zp}$ with the exception that $\langle
\tau^m \rangle$ no longer depends on power, $m$. As a result,
$\mathcal{C}_3(r = p^k)$ decays as a fractal area-law no matter
what value $p$ takes: \beqn && \mathcal{C}_3(r = p^k) =  \langle
\tau \rangle^{r^{d_H}}, \cr\cr &=& \bigg ( \frac{e^{(
\frac{p+1}{2}) \beta t} - e^{ \frac{\beta t}{2}}}{e^{(
\frac{p+1}{2}) \beta t} + (p-1) e^{ \frac{\beta t}{2}}} \bigg
)^{r^{d_H}} \sim  e^{-\alpha r^{d_H}}. \eeqn

The $Z_p$ Pascal's triangle model with prime integer $p$ can be
further extended to $Z_N$ models where $N$ is a composite positive
integer. These new composite $Z_N$ models have more than one
fractal symmetry. In fact, there is a distinct $Z_p$ fractal
symmetry for each unique prime divisor of $N$, e.g. the $Z_{N=6}$
model has a $Z_2$ Sierpinski-triangle fractal symmetry and a $Z_3$
Pascal's triangle modulo $3$ fractal symmetry. The behavior of the
three-point correlations of the $Z_N$ models is further discussed
in the SM.

{\it The Quantum Phase diagram:} So far we have only discussed the
classical version of the models. To turn quantum fluctuations on
in Eq.~\ref{U(1)}, one can modify the model as \beqn H_{Q-\U(1)} =
\sum_{\bigtriangledown} - t \cos\left( \theta_{1} + \theta_{2} +
\theta_{3}\right) + \sum_j \frac{U}{2} n_j^2, \label{qU(1)}\eeqn
where $n_j$ is the boson number operator defined on each site of
the triangular lattice, which are conjugate to the boson phase
$[n_i, \theta_j] = \ii \delta_{ij}$.

As shown previously, a class of ground states of the classical
$\U(1)$ model Eq.~\ref{U(1)} have the conventional
$\sqrt{3}\times\sqrt{3}$ order, which spontaneously breaks the
Pascal's triangle symmetry, and the two $\U(1)$ symmetries in
Eq.~\ref{2U(1)}. We now investigate whether this classical order
is stable against quantum fluctuation, i.e. whether it is stable
against the $U$ term in Eq.~\ref{qU(1)}. We argue that all
symmetries of the Hamiltonian (\ref{qU(1)}) are restored by
quantum fluctuations.

In an ordered phase that spontaneously breaks the $\U(1)$ global
symmetry, one can ignore the fact that the phase angle $\theta$ is
a compact boson (i.e. $\theta \sim \theta + 2\pi$), and hence
expand the cosine function of Eq.~\ref{qU(1)} to the lowest
nontrivial order. This procedure leads to an approximate Gaussian
Hamiltonian: \beqn H^{g}_{Q-\U(1)} = \sum_{i,j} t \theta_i
\theta_j + \sum_j 3 t\theta_j^2 + \frac{U}{2} n_j^2.
\label{qU(1)g} \eeqn The band structure of $\theta$ based on this
Gaussian Hamiltonian has minima at $\pm \mathbf{K} = \pm (4\pi/3,
0)$, which is consistent with the $\sqrt{3}\times\sqrt{3}$ order
of the classical Hamiltonian. The spectrum of the Gaussian
Hamiltonian is gapless.

The Gaussian expansion of the Hamiltonian ignored the compactness
of $\theta$. In a quantum model constructed with $\theta$,
$\theta$ being a compact boson is equivalent to the constraint
that its quantum conjugate variable $n$ take discrete values. To
check the stability of a semiclassical state of $\theta$ under
quantum fluctuation, one needs to investigate whether the
compactness of $\theta$, or equivalently the discrete nature of
$n$ would destabilize the semiclassical state described by the
Gaussian Hamiltonian Eq.~\ref{qU(1)g}. For example, the $(2+1)d$
quantum dimer model on the square lattice can be mapped to a
compact $\U(1)$ gauge theory~\cite{RK,dimerqed}, a Gaussian
expansion would lead to gapless photons. But the compactness of
the gauge field is always relevant in a semiclassical photon state
unless the system is at a fine-tuned multicritical point (the
so-called RK point~\cite{RK}), hence the Gaussian state is
generally unstable against quantum fluctuations. This effect is
also referred to as confinement of a lattice gauge theory. The
standard method of the analysis relies on the dual formalism of
Eq.~\ref{qU(1)} and Eq.~\ref{qU(1)g}. The dual model is defined on
the dual triangular lattice (Fig.~\ref{dualtriangle}), by
introducing the following variables: \beqn \sum_{j \in
\bigtriangledown} \theta_j = \phi_{\bar{j}}, \ \ \ \ n_ j =
\sum_{\bar{j} \in \triangle \ \mathrm{around} \ j} \Psi_{\bar{j}}.
\label{eq:lattice_duality}\eeqn where $\bar{j}$ denotes the sites
of the dual triangular lattice. $\phi_{\bar{j}}$ and
$\Psi_{\bar{j}}$ are canonically conjugate variables.
$\Psi_{\bar{j}}$ takes half-integer values, and $\phi_{\bar{j}}$
is compact.  The dual Hamiltonian reads \beqn H_{Q-\U(1)}^d =
\sum_{\bar{j}} - t\cos(\phi_{\bar{j}}) + \sum_{\bar{\triangle}}
\frac{U}{2} ( \Psi_{\bar{1}} + \Psi_{\bar{2}} + \Psi_{\bar{3}})^2.
\label{dual}\eeqn

Instead of directly dealing with the discrete variable $\Psi$, we
may view $\Psi_{\bar{j}}$ as taking continuous values, and
$\phi_{\bar{j}}$ as its non-compact conjugate variable. The
discrete nature of $\Psi$ can be enforced through an external
potential in the dual Hamiltonian. The dual Hamiltonian becomes
\beqn H_{Q-\U(1)}^d &\sim& \sum_{\bar{\triangle}} \frac{U}{2} (
\Psi_{\bar{1}} + \Psi_{\bar{2}} + \Psi_{\bar{3}})^2 -
\sum_{\bar{j}} t\cos(\phi_{\bar{j}}) \cr\cr &-& \alpha \cos(2\pi
\Psi_{\bar{j}}). \label{dual1} \eeqn The next step is to
temporarily ignore the $\alpha$-terms, and expand $ -
t\cos(\phi_j)$ to the lowest nontrivial order. After this
procedure, the dual Hamiltonian takes a Gaussian form, and it is
the dual of the Gaussian Hamiltonian Eq.~\ref{qU(1)g}. The goal of
this analysis is to check the role of the $\alpha$-terms at this
Gaussian state. This dual Gaussian Hamiltonian can be solved,
leading to a band structure of $\Psi$. The minima of the band
struture of $\Psi$ are located at the two corners of the Brillouin
zone, $\pm \mathbf{K} = (\pm 4\pi/3, 0)$. We then expand
$\Psi_{\mathbf{r}}$ at $\pm \mathbf{K}$: \beqn \Psi (\mathbf{r})
\sim e^{\ii \mathbf{K} \cdot \mathbf{r} } \psi(\mathbf{r}) + e^{-
\ii \mathbf{K} \cdot \mathbf{r} } \psi^\ast(\mathbf{r}). \eeqn The
Lagrangian of the dual theory expanded at $\pm \mathbf{K}$ becomes
\beqn \mathcal{L}_{Q-\U(1)}^d = (\partial_\tau \vec{\psi})^2 +
\rho_2 (\nabla \vec{\psi})^2 - \sum_{a}\alpha \cos(\vec{e}_a \cdot
\vec{\psi}), \label{dual2} \eeqn where $\vec{\psi} =
(\mathrm{Re}(\psi), \mathrm{Im}(\psi))$; $a = A, B, C$ label the
three sublattics of the dual triangular lattice, and $e_A = 2\pi
(1, 0)$, $e_B = 2\pi(-1/2, \sqrt{3}/2)$, $e_C = 2\pi(-1/2,
-\sqrt{3}/2)$.

The last three terms in Eq.~\ref{dual2} arise from rewriting the
last term of Eq.~\ref{dual1} by expanding $\Psi$ at $\pm
\mathbf{K}$. After this expansion, the last term of
Eq.~\ref{dual1} becomes $- \alpha \cos(\vec{e}_a \cdot
\vec{\psi}(\mathbf{r}))$ for $\mathbf{r}$ belonging to sublattice
$a$ ($a = A, B, C$) of the dual triangular lattice. Hence at long
scale a nonvanishing term would survive. The $\alpha$ term in
Eq.~\ref{dual2} will be relevant for the Gaussian theory with
nonzero $\rho_2$, which implies that the compactness of $\theta$,
or the discrete nature of $n$ in Eq.~\ref{qU(1)} destabilizes the
semiclassical Gaussian state, and the spectrum of Eq.~\ref{qU(1)}
should be gapped even with small $U$.

The dual description of the U(1) model studied here accurately
captures the spectrum of the gapless modes arising from
spontaneously breaking the global U(1) symmetries, though it does
not reproduce the spectrum at $U = 0$ that arise due to the Pascal
triangle symmetries.  Nevertheless, the nature of the ground-state
of the system when the pinning potential flows to strong coupling
can still be inferred. A strong $\alpha$ would pin $\Psi$ to
integer values, which implies that a relevant $\alpha$ would drive
the system into an eigenstate of $n$ in Eq.~\ref{qU(1)}, and the
$U$ term will lead to a unique and gapped ground state without any
spontaneous symmetry breaking. Hence we postulate that quantum
fluctuations of Eq.~\ref{qU(1)} restores all the symmetries of
model Eq.~\ref{U(1)}, and continuously connects to the large$-U$
limit of Eq.~\ref{qU(1)}. The analysis here would be more involved
if $n$ takes half integer values in Eq.~\ref{qU(1)}.

One possible quantum generalization of Eq.~\ref{Zp} is \beqn
H_{Q-Z_p} = \sum_\dtriangle -t \sigma_1^z \sigma_2^z \sigma_3^z  -
\sum_j h \sigma^x_j + \text{h.c.} \label{qZp} \eeqn for which the
clock operators $\sigma^z$ and $\sigma^x$ obey $(\sigma^z)^p =
(\sigma^x)^p = 1$ and $\sigma^z \sigma^x = e^{2 \pi i/p} \sigma^x
\sigma^z$ (one can also take $\sigma^z= \exp(\ii \theta)$ and
$\sigma^x = \exp(\ii 2\pi n / p)$, and restrict $\theta$ to take
values in $\frac{2 \pi}{p}Z_p$). Unlike the quantum $\U(1)$ model,
Eq. $\ref{qZp}$ is exactly self-dual with the introduction of the
dual plaquette variables \beqn \tau^x_{\bar{j}} = \sigma_1^z
\sigma_2^z \sigma_3^z, \ \ \ \ \tau^z_{\bar{1}} \tau^z_{\bar{2}}
\tau_{\bar{3}}^z = \sigma^x_j \eeqn for which the dual Hamiltonian
takes the same form as Eq.~\ref{qZp} with $t$ and $h$ switched.
Since the spectrum of the $Z_p$ models at $h = 0$ is gapped, and
it takes infinite order of perturbations of $h$ to mix two
different ground states in the thermodynamics limit, the classical
fractal-order of the $Z_p$ Pascal's triangle models is not
destroyed upon the introduction of quantum fluctuations.
Furthermore, the exact self-duality implies that there should be
one or more quantum phase transitions that separate the fractal
ordered phase $(t \gg h)$ and the disordered phase $(h \gg t)$.

{\it Discussion:} Although we demonstrated that the semiclassical
order of Eq.~\ref{qU(1)} is unstable against quantum fluctuation,
some deformation of Eq.~\ref{qU(1)} can support a stable
semiclassical order. In the SM we will show that if we sum over
three-boson interactions for both upward-facing and
downward-facing triangles, the semiclassical $\sqrt{3}\times
\sqrt{3}$ order becomes stable against quantum fluctuations. Also,
the system may be tuned to a multicritical point where $\rho_2$ in
Eq.~\ref{dual2} vanishes, and the low energy dynamics is
controlled by $\rho_4 (\nabla^2 \vec{\psi})^2$. The system can
remain gapless for a finite range of $\rho_4$, though it takes
tuning multiple parameters to reach this
state~\cite{dimer2,dimer3,dimer4}.

The nature of the quantum phase transition(s) in the quantum $Z_p$
model is a challenging subject. So far there is no well
established paradigm for understanding quantum phase transitions
involving spontaneous breaking of a fractal symmetry. Any approach
to study the quantum phase transition of the $Z_p$ models (such as
the quantum $Z_2$ Sierpinski-triangle model) through the $\U(1)$
generalization would need to address the enlarged Pascal's
triangle symmetry pointed out in the current work.

C.X. is supported by NSF Grant No. DMR-1920434, and the Simons
Investigator program.

\bibliography{fractal}

\appendix

\section{Construction of the $\U(1)$ parent model}

The three-body interaction of the $\U(1)$ parent model is
artificial because the nature is dominated by two-body
interactions. In what follows, we will discuss a more natural
construction of the $\U(1)$ model through a set-up with only
two-body interactions. The microscopic system we start with is a
honeycomb lattice, with a spin-3/2 degree of freedom $\vec{S}_R$
on each site $R$ of sublattice $\cA$ of the honeycomb lattice, and
a spin-1/2 degree of freedom $\vec{s}$ on each site of sublattice
$\cB$. The initial model only has two body interactions: \beqn H_0
&=& \sum_{R \in \cA} \sum_{b = 1}^3 \ J\left( S^x_{R} s^x_{R,b} +
S^y_{R} s^y_{R,b} \right)- D (S^z_R)^2 \cr\cr &=& \sum_{R \in \cA}
\sum_{b = 1}^3 \ \frac{J}{2} \left( S^+_{R} s^-_{R,b} + S^-_{R}
s^+_{R,b} \right)- D (S^z_R)^2 \label{H0} \eeqn The first sum of
$R$ is over all the sites on sublattice $\cA$, the second sum
$\sum_{b = 1}^3$ is over the three neighboring sites on sublattice
$\cB$ around $R$. We also consider other interactions between
second-neighbor sites, or equivalently first-neighbor sites within
respective sublattices $\cA$ and $\cB$: \beqn H' = \sum_{R,R' \in
\cA} J_1 (S^x_R S^x_{R'} + S^y_R S^y_{R'}) + \sum_{i,j \in \cB}
J_2 (s^x_i s^x_j + s^y_i s^y_j). \label{H'}\eeqn

\begin{figure}
\begin{center}
\includegraphics[width=0.45\textwidth]{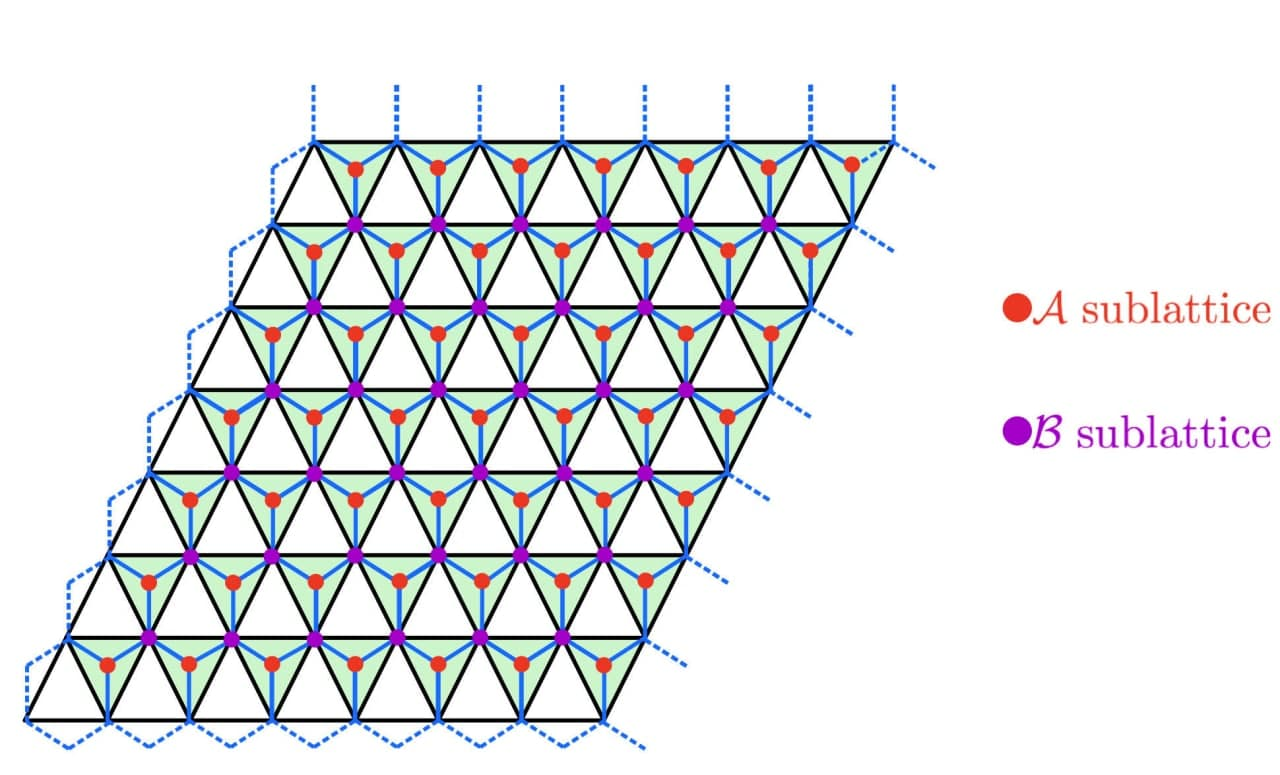}
\caption{The honeycomb lattice where Eq.~\ref{H0} is defined on.
The sublattice $\cB$ of the honeycomb lattice coincides with the
sites of the triangular lattice of Eq.~\ref{st} and
Eq.~\ref{U(1)}.}
\end{center}\label{honeycomb}
\end{figure}

The $D$ term in Eq.~\ref{H0} is an onsite spin anisotropy term. If
$D$ is positive and large, at low energy only the process that
tunnels between $S^z = -3/2$ and $S^z = +3/2$ is allowed. At the
third order perturbation of $J$, a nontrivial term is generated
within the low energy Hilbert space, which acts as a low energy
effective Hamiltonian: \beqn H_{\mathrm{eff}, 0} &=& \sum_{R \in
\cA} K \left( (S^+_R)^3 s^-_{R,1}s^-_{R,2}s^-_{R,3} + (S^-_R)^3
s^+_{R,1}s^+_{R,2}s^+_{R,3} \right), \cr\cr K &\sim&
\frac{J^3}{D^2}. \label{He0} \eeqn Notice that at low energy, only
terms with $(S^+)^3$ and $(S^-)^3$ can survive, while terms like
$(S^+_R)^3 (s^-_{R,1})^2 s^-_{R,2}$ must vanish because $(s^-)^2 =
0$ for spin-1/2 operator $s^{\pm}$.

Two other effects need to be discussed: (1) at the second order
perturbation of $J/D$, the $J_2$ term in Eq.~\ref{H'} will be
renormalized, and shifted by $\sim J^2/D$. We assume that the
renormalized $J_2$ is very small and negligible. (2) The 3rd order
perturbation of $J_1$ will generate a new term \beqn
H_{\mathrm{eff}}' \sim \sum_{R,R' \in \cA} \frac{J_1^3}{D^2}
\left( (S_R^+)^3 (S_{R'}^-)^3 + \text{h.c.} \right) \label{He'}
\eeqn Now, if we combine Eq.~\ref{He0} and Eq.~\ref{He'} together
and use the standard spin-boson mapping: $S^+_R \sim \exp(\ii
\theta_{R})$ and $s^+_{R,b} \sim \exp(\ii \theta_{R,b})$, the
entire low energy effective Hamiltonian is mapped to \beqn
H_{\mathrm{eff}} &=& \sum_{R \in \cA} - t \cos\left( 3 \theta_R -
\theta_{R,1} - \theta_{R,2} - \theta_{R,3}\right) \cr\cr &-&
\sum_{R,R' \in \cA} t' \cos(3\theta_R - 3\theta_{R'}).
\label{He}\eeqn Due to the $t'$ term, at zero temperature
$3\theta_R$ will order on $\cA$ sublattice. If we replace
$3\theta_R$ with its expectation value, then Eq.~\ref{He} reduces
to Eq.~\ref{U(1)} on a triangular lattice, which corresponds to
the sublattice $\cB$ of the original honeycomb lattice.

When we map the spin model to a boson rotor model, there is one
extra subtlety that the boson number takes half-integer, rather
than integer values. This to some extent complicates the analysis
of the phase where $\theta$ is disordered, due to the extra
degeneracy arising from $S^z$.

\section{Pascal's Triangle Symmetry of the $\U(1)$ Parent Model}\label{app:Pascal_Symm}
In this section, we establish the presence of a generalized
fractal symmetry of the $\U(1)$ parent Hamiltonian
(\ref{eq:three_boson}) given an appropriate set of boundary
conditions.  When these boundary conditions are not satisfied, we
establish that these approximate symmetries give rise to
low-energy modes.  Furthermore, we show that the $Z_{p}$
generalization of the Sierpinski triangle model exhibits a
ground-state degeneracy of $D = p^{L-1}$ on an $L\times L$ system
with periodic boundary conditions, when $L = p^{k}-1$.

\begin{figure}[t]
\includegraphics[width=0.3\textwidth]{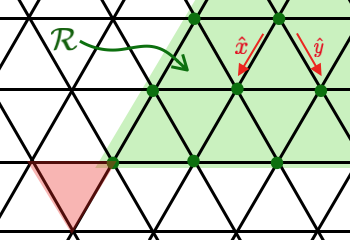}
\caption{No excitations are created at the upward-facing triangles
that border an edge of region $\mathcal{R}$, as shown.}
\label{app:bottom_row_R}
\end{figure}

For the $\U(1)$ parent Hamiltonian, starting with the
lowest-energy configuration with $\theta_{i} = 0$ everywhere (with
energy $E_{0}$), we may perform the transformation
\begin{align}\label{app:theta_transf}
     \theta_{i} \rightarrow \theta_{i} + \frac{2\pi}{p} (-1)^{i_{x}+i_{y}}\left(\begin{array}{c}i_{x}+i_{y}\\ i_{y} \end{array}\right)
\end{align}
within a region $\mathcal{R}_{(0,0)}$ which consists of the points
contained within an equilateral triangle of side-length $p^{k}$
with the top of the triangle at the origin $(i_{x},i_{y}) =
(0,0)$, i.e.
\begin{align}
    \mathcal{R}_{(0,0)} \equiv \left\{ (i_{x},i_{y}) \,|\, 0\le i_{y}\le i_{x}\,\mathrm{and}\, i_{x}+i_{y}\in[0,p^{k}-1]\right\}\nonumber
\end{align}
where $p$ is a {\it prime} number and $k\ge 0$ is an integer.  This transformation creates three excitations -- downward-facing triangles at which $\theta_{i} + \theta_{i+\hat{x}} + \theta_{i+\hat{x}-\hat{y}} \ne 0$ mod $2\pi$ -- at each corner of the triangular region $\mathcal{R}$.  The energy of the final state is
\begin{align}
    E_{0} + 3t\left[1 - \cos\left(\frac{2\pi}{p}\right)\right]\label{app:energy_exc_state}
\end{align}.

We may demonstrate this as follows. The transformation (\ref{app:theta_transf}) creates an excitation at the downward-facing triangle whose bottom corner sits at the origin  since $\theta_{-\hat{x}}+\theta_{0}+\theta_{-\hat{y}} \rightarrow \theta_{-\hat{x}}+\theta_{0}+\theta_{-\hat{y}} + (2\pi/p)$ under the transformation (\ref{app:theta_transf}). Identical excitations are created at the two downward-facing triangles at the two other corners of $\mathcal{R}_{(0,0)}$.

We now demonstrate that the transformation (\ref{app:theta_transf}) acts trivially everywhere away from the corners of $\mathcal{R}_{(0,0)}$.  First, this transformation does not create any excitations at downward-facing triangles that are contained entirely within the region $\mathcal{R}_{(0,0)}$, i.e. $\theta_{i} + \theta_{i+\hat{x}} + \theta_{i+\hat{x}-\hat{y}}$ is unchanged by the transformation for any triangle for which $i$, $i+\hat{x}$, and $i+\hat{x}-\hat{y}$ are contained entirely in $\mathcal{R}_{(0,0)}$. This can be verified explicitly by applying the transformation (\ref{app:theta_transf}).

Now, we may consider downward-facing triangles that border an edge of $\mathcal{R}_{(0,0)}$.  Consider downward-facing triangles that border the bottom of region $\mathcal{R}_{(0,0)}$, as shown in Fig. \ref{app:bottom_row_R}. The vertices of these triangles are located at points $(i,i+\hat{x},i+\hat{x}-\hat{y})$ for which $i$, $i+\hat{x}-\hat{y}\in\mathcal{R}$ while $i+\hat{x}\notin\mathcal{R}$. It can be seen that for these downward facing triangles, the transformation (\ref{app:theta_transf}) leads to the shift
\begin{widetext}
\begin{align}
    \theta_{i} + \theta_{i+\hat{x}} + \theta_{i+\hat{x}- \hat{y}}
    &\rightarrow \theta_{i} + \theta_{i+\hat{x}} + \theta_{i+\hat{x}- \hat{y}} +  \frac{2\pi}{p}(-1)^{i_{x}+i_{y}}\left[\left(\begin{array}{c} i_{x}+i_{y}\\i_{y}\end{array}\right) + \left(\begin{array}{c} i_{x}+i_{y}\\i_{y}-1\end{array}\right)\right] = \theta_{i} + \theta_{i+\hat{x}} + \theta_{i+\hat{x}-\hat{y}} +  \frac{2\pi}{p}\left(\begin{array}{c}p^{k}\\i_{y}\end{array}\right)\nonumber
\end{align}
\end{widetext}
In the last equality, we've used the fact that $i_{x}+i_{y} = p^{k}-1$ for the downward-facing triangles bordering the bottom edge of $\mathcal{R}_{(0,0)}$. We now observe that
\begin{align}
    \left(\begin{array}{c}p^{k}\\i_{y}\end{array}\right) = 0\mod\hspace{.05in} p
\end{align}
for any $1\le i_{y}\le p^{k}-1$. This may be proven by observing that
\begin{align}
    i_{y}\left(\begin{array}{c}p^{k}\\i_{y}\end{array}\right) = p^{k}\left(\begin{array}{c}p^{k}-1\\i_{y}-1\end{array}\right)\label{app:prime_mod_0}
\end{align}
The prime $p$ divides the right-hand-side of (\ref{app:prime_mod_0}) at least $k$ times. Furthermore, since $1\le i_{y}\le p^{k}-1$, $p$ divides $i_{y}$ at most $k-1$ times. As a result, $p$ must divide $\left(\begin{array}{c}p^{k}\\i_{y}\end{array}\right)$ at least once, which completes the proof.  From this, we conclude that $\theta_{i} + \theta_{i+\hat{x}} + \theta_{i+\hat{x}- \hat{y}}$ is shifted by an integer multiple of $2\pi$ for downward-facing triangles that border the bottom edge of $\mathcal{R}_{(0,0)}$.  Therefore, when starting from the state with $\theta_{i} = 0$ everywhere, only three excitations are created by the transformation (\ref{app:theta_transf}) which lie at the corners of $\mathcal{R}_{(0,0)}$, and the energy cost of the final state is indeed given by (\ref{app:energy_exc_state}).

We conclude by determining the ground-state degeneracy of the $Z_{p}$ models on an $L\times L$ triangular lattice with periodic boundary conditions, for particular values of $L$.  First, the previous discussion of the Pascal's symmetry implies that if we start from a configuration with $\theta_{i} = 0$ everywhere and perform a symmetry transformation over the region $\mathcal{R}_{(0,0)}$, then the configuration of $\theta_{i}$ along the row defined by $i_{x}+i_{y} = p^{k}-1$ of the lattice is given by \begin{align}
    \theta_{i} \,\,\mathrm{mod}\,\,2\pi= \left\{\begin{array}{cc}
    \displaystyle(-1)^{i_{y}}\frac{2\pi}{p}  & \,\,\,\,\,\,i_{y}\in [0,p^{k}-1]\\
    & \\
    0 & \mathrm{otherwise}
    \end{array}\right.\nonumber
\end{align}
Now consider another symmetry transformation along the region $\mathcal{R}_{(-1,1)}$, i.e. along the equilateral triangle of side length $p^{k}$ whose top vertex lies at the site $(i_{x},i_{y}) = (-1,1)$.  Superposing this transformation with the previous transformation along $\mathcal{R}_{(0,0)}$, we find that along the row $i_{x}+i_{y} = p^{k}-1$, that
\begin{align}
    \theta_{i} \,\,\mathrm{mod}\,\,2\pi= \left\{\begin{array}{cc}
    \displaystyle\frac{2\pi}{p}  & \,\,\,\,\,i_{y} = 0\,\,\mathrm{or}\,\, p^{k}\\
    & \\
    0 & \mathrm{otherwise}
    \end{array}\right.\label{eq:theta_transf}
\end{align}
If we now impose periodic boundary conditions so that $(i_{x},i_{y})\sim (i_{x}+L,i_{y})\sim(i_{x},i_{y}+L)$ where $L = p^{k}-1$, we observe that the symmetry transformations along $\mathcal{R}_{(0,0)}$ and $\mathcal{R}_{(-1,1)}$ have generated a new ground-state of the system, since the row $i_{x}+i_{y} = p^{k}-1$ is now identified with row $i_{x} + i_{y} = 0$, and the configuration (\ref{eq:theta_transf}) is precisely the configuration of $\theta_{i}$ along the first row $i_{x}+i_{y} = 0$.

By applying additional ``pairs" of these symmetry transformations (e.g. along $\mathcal{R}_{(i_{x},-i_{x})}$ and $\mathcal{R}_{(i'_{x},-i'_{x})}$), we may generate other ground-state configurations.  Each unique ground-state corresponds to a unique configuration of $\theta_{i} = \frac{2\pi}{p}m_{i}$ for $m_{i}\in [0,\ldots,p-1]$ at each site  along the row $i_{x}+i_{y} = 0$.  Not all of these $p^{L}$ configurations can be reached by starting with the ground-state with $\theta_{i}=0$ everywhere and applying the fractal symmetry transformation in pairs.  It is clear, however, that by continuing to apply these symmetry transformations in pairs, any configuration of $\theta_{i}$ along the row $i_{x}+i_{y} = 0$ can be reached for which
\begin{align}
    \sum_{i|i_{x}+i_{y}=0}(-1)^{i_{y}}m_{i}=0\hspace{.1in}\mathrm{mod}\,\,p
\end{align}
There are precisely $p^{L-1}$ such configurations.  We conclude that the ground-state degeneracy of the $Z_{p}$ model is $p^{L-1}$ on an $L\times L$ lattice with periodic boundary conditions when $L = p^{k}-1$.

\section{Three-point correlation of the $Z_p$ and $Z_N$ models}
Starting from the classical $Z_p$ Pascal's triangle models in Eq. $\ref{Zp}$,
since the $Z_p$ clock degrees of freedom are primitive
$p$-roots of unity for which $\sigma^\dagger = \sigma^{p-n}$, $\Re{\sigma} = \Re{\sigma^{p-n}}$, $\Im{\sigma} = -
\Im{\sigma^{p-n}}$ and likewise for the dual $\tau$-variables. As a result, for the partition function of a single $\tau$-variable $Z_1 = \sum_{n=0}^{p-1} e^{\beta t \cos( \frac{2 \pi n}{p} )}$: \beqn
\langle \Im{\tau^m} \rangle &=& \frac{1}{Z_1} \sum_{n=0}^{p-1}
\sin \bigg ( \frac{2 \pi m n}{p} \bigg ) e^{\beta t \cos ( \frac{2
\pi n}{p} )},
\cr\cr &=& 0;\\
\langle \Re{\tau^m} \rangle &=& \frac{1}{Z_1} \sum_{n=0}^{p-1}
\cos \bigg ( \frac{2 \pi m n}{p} \bigg ) e^{\beta t \cos ( \frac{2
\pi n}{p} )}, \cr\cr &=& \frac{1}{Z_1} \sum_{n=0}^{p-1} \cos \bigg
( \frac{2 \pi m (p-n)}{p} \bigg ) e^{\beta t \cos ( \frac{2 \pi
n}{p} )}, \cr\cr &=& \langle \Re{(\tau^m)^\dagger} \rangle. \eeqn
As a consequence, $\langle \tau^m \rangle$ = $\langle
(\tau^m)^\dagger \rangle$. In the dual representation, the
three-point correlation function factors since each
$\tau$ is statistically independent \beqn
\mathcal{C}_3(r = p^k) &=& \langle \sigma_{0,0} \sigma_{r,0}
\sigma_{0,r} \rangle, \cr\cr &=& \langle \text{product of }\tau^m
\text{ in Pascal's triangle} \rangle, \cr\cr &=& \text{product of }
\langle \tau^m \rangle \text{ in Pascal's triangle}, \cr\cr &=&
\prod_{m=1}^{\frac{p-1}{2}} \langle \tau^m \rangle^{N_{m,p-m}(k)} \label{3pointder},
\eeqn where $N_{m,p-m}(k)$ is the number
of times $m$ and $p-m$ appear in the Pascal's triangle modulo $p$ of
length $p^{k}-1$. Note that the staggering of $\tau$-variables in the representation of $\sigma$ is inconsequential for the calculation of the three-point correlation as $\langle \tau^m \rangle = \langle (\tau^m)^\dagger \rangle$.

As a result, computing the three-point correlation function
reduces to counting the number of times each unique entry in a Pascal's triangle modulo $p$ appears. Since the
elements of a Pascal's triangle modulo $p$ are binomial coefficients
${j \choose i} \mod p$, one may in principle calculate
$N_{m,p-m}(k)$ by brute force according to \beqn N_{m,p-m}(k) &=&
\sum_{i \leq j} \sum_{j=0}^{p^k-1} \delta \bigg ( m, {j \choose
i}\text{ mod }p \bigg ) \cr\cr &+&\sum_{i \leq j}
\sum_{j=0}^{p^k-1} \delta \bigg ( p-m, {j \choose i}\text{ mod } p
\bigg ) \label{bruteforce}. \eeqn However, this is neither
efficient analytically nor numerically, even at small $p$. It is
instead desirable to derive recurrence relations that can be
solved assuming initial conditions $N_{m,p-m}(1)$ are provided
which can be computed according to Eq. $\ref{bruteforce}$ with $k
= 1$. In order to derive such recurrence relations, we can make
use that for $\cos(m \phi), m = 1,\dots,
\frac{p-1}{2}$ such that \beqn \cos(m \phi) = \cos(qn \phi)
\Leftrightarrow
\begin{cases}
m = qn \text{ mod }p\\
m = q(p - n) \text{ mod }p
\end{cases} \label{cyclic}.
\eeqn This implies that $\langle \tau^m \rangle = \langle
\tau^{qn} \rangle$ where $q = q(m,n)$ is the smallest element of $Z_p$
for which one of the two equations in Eq. $\ref{cyclic}$ is
solved. If we consider an arbitrary Pascal's triangle modulo $p$ of
length $p^k-1$, this triangle is self-similar and can be
constructed by tiling the same Pascal's triangle of length
$p^{k-1}-1$ according to the Pascal's triangle of length $p-1$. The
number of $\langle \tau^n \rangle$ which become $\langle \tau^{m}
\rangle$ upon tiling is $N_{q(m,n),p-q(m,n)}(1)$. As a result, we
arrive at a set of $\frac{p-1}{2}$ coupled recurrence relations
\beqn N_{m,p-m}(k) = \sum_{n=1}^{\frac{p-1}{2}}
N_{q(m,n),p-q(m,n)}(1) N_{n,p-n}(k-1) \label{recurrence} \eeqn

For small $p$, $N_{m,p-m}(1)$ can be calculated and these
recurrence relations can be solved without the use of numerics.
For example, for the $p = 5$ model
\begin{align*}
q =
\begin{bmatrix}
1&2\\
2&1
\end{bmatrix}.
\end{align*}
One can compute the initial conditions by hand from Eq. $\ref{bruteforce}$ to find $N_{1,4}(1) = 12$ and $N_{2,3}(1) = 3$, yielding coupled recurrence relations
\beqn
N_{1,4}(k) &=& 12N_{1,4}(k-1)+3N_{2,3}(k-1),\\
N_{2,3}(k) &=& 12N_{2,3}(k-1)+3N_{1,4}(k-1).
\eeqn
This can be solved again using the initial conditions to recover
\begin{align*}
N_{1,4}(k) = \frac{3^{k-1}}{10}(3 \times 5^{k+1} + 5 \times 3^{k+1}),\\
N_{2,3}(k) = \frac{3^{k-1}}{10}(3 \times 5^{k+1} - 5 \times
3^{k+1}).
\end{align*}
One can verify that $N_{1,4}(k)+N_{2,3}(k)$ must be the area of the fractal of length $p^k-1$, $N_{1,4}(k)+N_{2,3}(k) = 15^k = 5^{k \times
\frac{\ln 15}{\ln 5}} = 5^{k \times d_H}$. The
three-point correlation can now be computed as \beqn
\mathcal{C}_3(r = 5^k) &=& \langle \tau \rangle^{N_{1,4}(k)}
\langle \tau^2 \rangle^{N_{2,3}(k)},
\cr\cr &=& \bigg ( \frac{1}{Z_1} \sum_{n=0}^{p-1} \cos \bigg ( \frac{2 \pi n}{p} \bigg ) e^{\beta t \cos(\frac{2 \pi n}{p} )} \bigg )^{N_{1,4}(k)} \cr\cr &\times&\bigg ( \frac{1}{Z_1} \sum_{n=0}^{p-1} \cos \bigg( \frac{4 \pi n}{p}  \bigg) e^{\beta t \cos(\frac{2 \pi n}{p} )} \bigg )^{N_{2,3}(k)}
\cr\cr &\approx& \bigg (
\frac{e^{\beta t} + 0.62e^{0.31 \beta t} - 1.62 e^{-0.81 \beta
t}}{e^{\beta t} + 2e^{0.31 \beta t} + 2e^{-0.81 \beta t}} \bigg
)^{N_{1,4}(k)} \cr\cr &\times & \bigg ( \frac{e^{\beta t} -
1.62e^{0.31 \beta t} + 0.62 e^{-0.81 \beta t}}{e^{\beta t} +
2e^{0.31 \beta t} + 2e^{-0.81 \beta t}} \bigg )^{N_{2,3}(k)}. \eeqn For sufficiently large distances, $\mathcal{C}_3(r = 5^k) \sim e^{-\alpha' r^\gamma}$ where $\gamma$ is an exponent that is larger than $d_H$, due to the decay of $\langle \tau^2 \rangle$ being stronger than $\langle \tau \rangle$. As a comparison, the three-point correlation in the $Z_5$ model decays much faster than the three-point correlation of the Sierpinski-triangle model and the $Z_3$ Pascal's triangle model.

The multiple fractal symmetries of the $Z_N$ models has important
consequences for the behavior of the three-point correlations as
well. All of the $Z_N$ models retain the same dual description,
and $\sigma$ can be represented as an infinite staggered product
of dual $\tau$-variables on a Pascal's triangle modulo $N$ in the
thermodynamic limit. However, since $N$ is composite, this
Pascal's triangle is not a fractal and hence there is no product
of three $\sigma$'s that has compact support causing
$\mathcal{C}_3(r)$ to vanish. Instead, for each unique prime
divisor $p$ of $N$ for which $q p = N$, $\sigma^{q}$ can be
represented as an infinite staggered product of $\tau^{q}$ in the
shape of a Pascal's triangle modulo $p$. Therefore,
$\mathcal{C}_3^{q}(r) = \langle \sigma^{q}_{0,0} \sigma^{q}_{r,0}
\sigma^{q}_{0,r} \rangle$ is non-vanishing when $r = p^k$. Using
$Z_{N=6}$ as an example again, $\mathcal{C}^{3}_3 = \langle
\sigma^3_{0,0} \sigma^3_{r,0} \sigma_{0,r}^3 \rangle$
characterizes SSB of the $Z_2$ fractal symmetry while
$\mathcal{C}^{2}_3 = \langle \sigma^2_{0,0} \sigma^2_{r,0}
\sigma_{0,r}^2 \rangle$ characterizes SSB of the $Z_3$ fractal
symmetry. Note that if $N$ is taken to be prime, since the only
prime-factor of $N$ is itself, these results reduce down to those
of the $Z_p$ models and $\langle \sigma_{0,0} \sigma_{r,0}
\sigma_{0,r} \rangle$ is ordered.

\begin{figure}
\begin{center}
\includegraphics[width=0.5\textwidth]{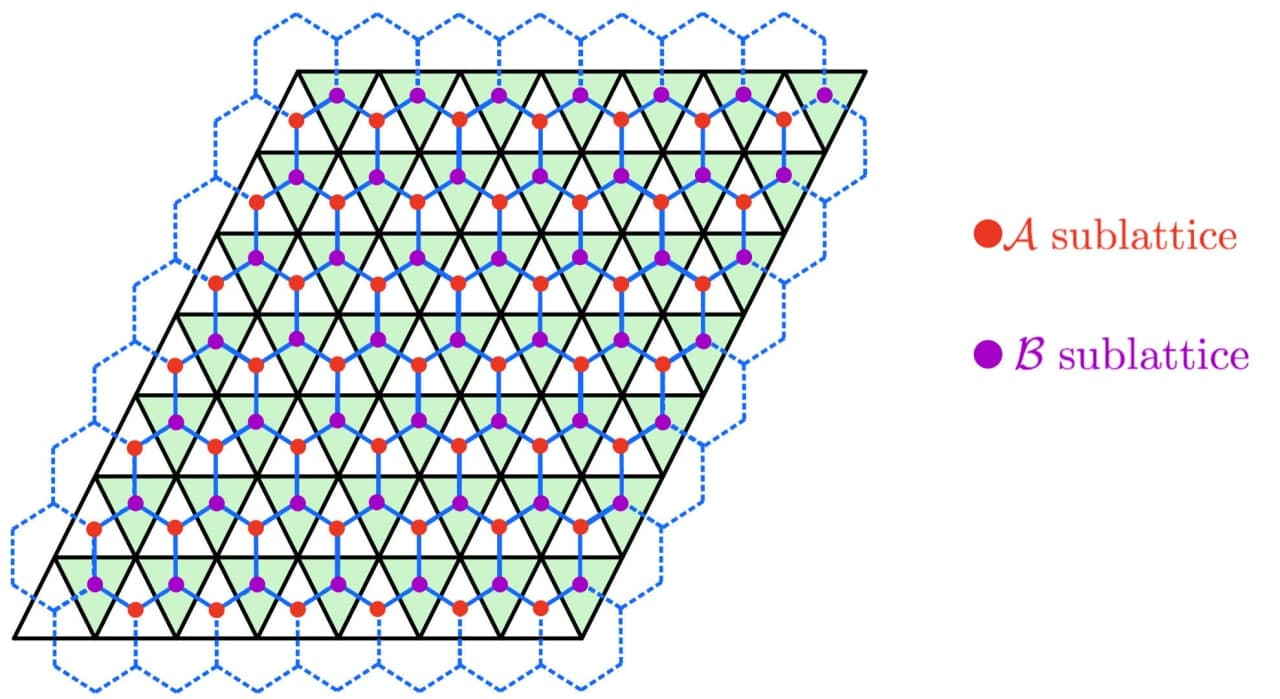}
\caption{The dual honeycomb lattice where Eq.~$\ref{qU(1)Bd}$ is
defined on. Each vertex of the honeycomb lattice coincides with
either a downward or upward facing plaquette of the original
triangular lattice.}\label{dualhoneycomb}
\end{center}
\end{figure}

\section{A related model with stable $\sqrt{3} \times \sqrt{3}$ order}

We consider a model directly related to Eq.~\ref{qU(1)}, but
instead is stable against the inclusion of quantum fluctuations.
In the $\U(1)$ parent model, the Pascal's triangle fractal
symmetries only exist because a single orientation of triangular
plaquettes is being summed over. If we include the sum over
three-boson interactions on upward facing plaquettes as well, the
model becomes \beqn H = \sum_{\dtriangle,\triangle} - t \cos\left(
\theta_{1} + \theta_{2} + \theta_{3}\right) + \sum_j \frac{U}{2}
n_j^2. \label{qU(1)B} \eeqn While this model does not have the
Pascal's triangle fractal symmetries of Eq.~\ref{qU(1)}, it still
retains the global $\U(1)_1 \times \U(1)_2$ symmetry in
Eq.~\ref{2U(1)} and hence the ground state of this model at small
$U$ will still have the $\sqrt{3}\times \sqrt{3}$ order. In this
case, unlike Eq.~\ref{qU(1)}, the boson ordered phase will
actually be stable against quantum fluctuations. This can be seen
by analyzing the dual of Eq.~\ref{qU(1)B}. Now, the dual variables
are defined on the sites $\bar{i}$ of the dual honeycomb lattice
(Fig.~\ref{dualhoneycomb}): \beqn && \theta_1 + \theta_2 +
\theta_3 = \phi_{\bar{i}}, \ \ \ \mathrm{site} \ 1, 2, 3 \
\mathrm{around} \ \bar{i}; \cr\cr && n_j =  \sum_{\bar{i}}
\Psi_{\bar{i}}, \ \ \ \bar{i} = 1, \cdots 6 \ \mathrm{around} \ j.
\eeqn The dual variable is subject to a local gauge constraint
around each site $j$: \beqn \sum_{\bar{i} \ \mathrm{around} \ j}
(-1)^{\bar{i}} \phi_{\bar{i}} = 0 \ \mathrm{mod} \ 2\pi. \eeqn
Again, $\phi_{\bar{i}}$ are compact variables, while
$\Psi_{\bar{i}}$ takes discrete values. The dual Hamiltonian of
Eq.~\ref{qU(1)B} reads~\cite{xuashvin} \beqn H_d = \sum_{\bar{j}}
- t \cos(\phi_{\bar{j}}) + \sum_{j} \frac{U}{2} \left(
\sum_{\bar{i} \ \mathrm{around} \ j} \Psi_{\bar{i}} \right)^2.
\label{qU(1)Bd} \eeqn The key difference from the model considered
in the main text is that, this dual model Eq.~\ref{qU(1)Bd} has a
gauge symmetry, i.e. the dual Hamiltonian is invariant under the
following local gauge transformation: \beqn \Psi_{\bar{i}}
\rightarrow \Psi_{\bar{i}} + (-1)^{\bar{i}} \epsilon_j, \ \ \
\bar{i} \ \mathrm{around} \ j, \label{gauge}\eeqn where
$\epsilon_j$ is an arbitrary integer-valued function on the
original triangular lattice $j$.

Now we can again replace $\Psi_{\bar{i}}$ by continuous variables,
and the constraint that $\Psi$ takes discrete values will be
enforced through ``vertex operators" such as $\cos(2\pi \Psi)$. If
we ignore these vertex operators, the Lagrangian of the Gaussian
theory of Eq.~\ref{qU(1)Bd} reads: \beqn \mathcal{L}_{d} =
\sum_{\bar{i}} \frac{1}{2t}(\partial_\tau \Psi_{\bar{i}}) +
\sum_{j} \frac{U}{2} \left( \sum_{\bar{i} \ \mathrm{around} \ j}
\Psi_{\bar{i}} \right)^2. \label{LQU(1)Bd} \eeqn This Gaussian
theory has a continuous gauge symmetry, which corresponds to
allowing $\epsilon_j$ to take arbitrary continuous values in
Eq.~\ref{gauge}. Then, the Gaussian fixed point of
Eq.~\ref{LQU(1)Bd} is stable against weak vertex operators, since
the vertex operators break the continuous gauge symmetry of the
Gaussian theory Eq.~\ref{LQU(1)Bd}.

\end{document}